\documentclass[12pt]{iopart}
\usepackage{graphics,amssymb}

\begin{document}

\title[Entropic sampling dynamics of the globally-coupled kinetic Ising model]
{Entropic sampling dynamics of the globally-coupled kinetic Ising model}

\author{Beom Jun Kim$^1$ and M Y Choi$^{2,3}$}

\address{$^1$ Department of Molecular Science
     and Technology, Ajou University, Suwon 442-749, Korea}
\address{$^2$ Korea Institute for Advanced Study, Seoul 130-722, Korea}
\address{$^3$ Department of Physics, Seoul National University,
Seoul 151-747, Korea}
\ead{beomjun@ajou.ac.kr}

\begin{abstract}

The entropic sampling dynamics based on the reversible information transfer
to and from the environment is applied to the globally coupled Ising model
in the presence of an oscillating magnetic field. 
When the driving frequency is low enough, coherence between the 
magnetization and the external magnetic field is observed; 
such behavior tends to weaken with the system size. The time-scale matching between the intrinsic time scale, 
defined in the absence of the external magnetic field, and the extrinsic time scale, 
given by the inverse of the driving frequency, is used to explain the observed
coherence behavior.

\end{abstract}


\maketitle

\section{Introduction}
The stochastic resonance~\cite{gammaitoni}, which has been extensively studied in various 
systems including many-body
systems~\cite{lindner,neda,leung,bjkimJJL,bjkim,Hong,kuperman,tome90,korniss}, 
refers to the phenomenon that an appropriate amount of stochastic noise may
not hinder but trigger the coherence between the output signal and the weak
periodic input signal. In Ref.~\cite{korniss}, a detailed study of the two-dimensional 
kinetic Ising model has been made and the size-dependence of the dynamic
phase diagram has been identified.
The globally-coupled kinetic Ising model in the presence of
a time-periodic external magnetic field has been investigated much, mainly because the
simple mean-field method can be applied~\cite{leung,tome90}. 
In particular, the 
model has been recently shown to possess double stochastic resonance peaks at two
distinct temperatures~\cite{bjkim}.  However, except for some
systems~\cite{glauber}, general applicability of the Glauber dynamics
is hardly justified in many real systems. 

The importance of the role of the entropy has been pointed out
in the biological evolution~\cite{mychoi}: In general, every species attempts to minimize 
its entropy, or in other words, to get negative entropy from the environment. 
From this idea, it has been shown that the biological evolution of an ecosystem can be well 
described by the so-called entropic sampling dynamics~\cite{jlee}.
We believe that the Ising model studied in this work can be related
to a wide range of phenomena in biological and social systems, like
the opinion formation problem in social systems under the time-periodic 
influence~\cite{kuperman}.
Very recently, it has been suggested that 
the ubiquitously observed self-organized critical behavior in nature 
is caused by the reversible information transfer with the surroundings
of the system~\cite{soc:mychoi}.  This implies that the entropic
sampling dynamics, which is based on the reversible information
exchange with the environment, can have realizations in many systems.


\section{Entropic Sampling Dynamics of Globally Coupled Ising Model}
We in this work use the globally coupled Ising model, the Hamiltonian
of which is given by
\begin{equation} \label{eq:H}
{\cal H} = -\frac{J}{2N}\sum_{i \neq j} \sigma_i \sigma_j - h\sum_i \sigma_i,
\end{equation}
where $\sigma_i = \pm 1$ is the Ising spin at site $i$, $J$ is the
coupling strength, $N$ is the total number of spins, and
the external magnetic field $h$ is coupled to
the total magnetization $M \equiv \sum_i \sigma_i$.
We first compute the entropy $S$ by using the entropic sampling
algorithm~\cite{jlee,wang-landau}, 
applied to the Hamiltonian in Eq.~(\ref{eq:H}) at constant magnetic field $h$.
The detailed balance condition for the entropic sampling reads~\cite{jlee}
\begin{equation} \label{eq:W}
\frac{W({\vec\sigma}\rightarrow {\vec\sigma}^\prime)}{W({\vec\sigma}^\prime\rightarrow{\vec\sigma})}
= e^{ -\{ S [E({\vec \sigma}^\prime)]  - S [E({\vec \sigma})] \} }, 
\end{equation}
where $W( {\vec \sigma} \rightarrow {\vec \sigma}^\prime) $ is the transition probability
from ${\vec \sigma} \equiv \{ \sigma_1, \sigma_2, \cdots, \sigma_N \} $ to 
${\vec \sigma}^\prime \equiv \{\sigma_1^\prime, \sigma_2^\prime, \cdots, \sigma_N^\prime \}$, and
for convenience, only a single spin is allowed to change at a given time.
In the entropic sampling algorithm, one initially starts from $S(E) = 0$ for all values 
of the energy $E$, and obtains the histogram $H(E)$ for several Monte Carlo sweeps, which
is then used to estimate a new value of $S(E)$:
\begin{equation}
S(E) = \left\{
\begin{array}{ll}
S(E) & \mbox{for $H(E) = 0$, } \\
S(E) + \ln H(E)  & \mbox{otherwise.} \\
\end{array}
\right.
\end{equation}
As the above procedure proceeds, $S(E)$ approaches the true entropy up to an additive constant, 
which is independent of the energy $E$ and thus may be subtracted on the condition 
that the minimum of $S(E)$ is zero.

Once the correct entropy is obtained, the time evolution under the
entropic sampling should satisfy the detailed balance condition in Eq.~(\ref{eq:W}) 
and we choose the following procedure:
(1) Generate ${\vec \sigma}^\prime$ which differs from ${\vec \sigma}$ only at
one spin, e.g., ${\vec \sigma}^\prime = \{\sigma_1, \sigma_2, \cdots , -\sigma_j, \cdots, \sigma_N \}$
obtained from ${\vec \sigma} = \{\sigma_1, \sigma_2, \cdots , \sigma_j, \cdots, \sigma_N \}$.
(2) Compute the entropy change 
$\Delta S \equiv S[E({\vec \sigma}^\prime)]  - S[E({\vec \sigma})]$.
(3) If $\Delta S \leq 0$, accept the try, i.e., change ${\vec \sigma}$ to ${\vec \sigma}^\prime$;
otherwise, accept the try with the probability $e^{-\Delta S}$.
One sweep of the above procedure for all the spins in the system corresponds
to one time unit in the present work.

\begin{figure}
\centering{\resizebox*{!}{6.0cm}{\includegraphics{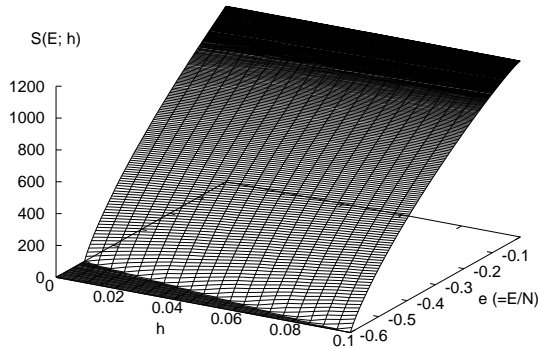}}}
\caption{Entropy $S(E;h)$ as a function of the energy $e \equiv E/N$ per spin
and the external magnetic field $h$. Each curve at a given value
of $h$ was obtained from the entropic sampling algorithm applied
to the Hamiltonian in Eq.~(\ref{eq:H}).
}
\label{fig:3d}
\end{figure}

Figure~\ref{fig:3d} shows the entropy $S(E;h)$, computed for the system of size
$N = 1600$ under time-independent field $h$, as a function of the energy per
spin, $e \equiv E/N$ (henceforth, we measure the energy in units of the
coupling strength $J$) obtained from $10^8$ sweeps per spin.  In the ground state all spins have the same value:
$\sigma_i = 1$ if $h>0$, and $\sigma_i = -1$ otherwise.  Accordingly, the
ground state energy is given by $E_{\rm ground} = -N/2 - hN$ or $e_{\rm ground}
\equiv E_{\rm ground}/N = -1/2 - h$.  It is observed in Fig.~\ref{fig:3d} that
$S(E;h) = 0$ for $e < e_{\rm ground}$.   One can also compute $S(E;h)$ 
analytically as follows: From Stirling's series expansion, it is
straightforward to get the entropy $S(m \equiv M/N)$ 
as a function of the magnetization,
\[
S(m) \approx -N \left[ 
\frac{1-m}{2}\ln\frac{1-m}{2} + 
\frac{1+m}{2}\ln\frac{1+m}{2} \right] .
\]
The entropy $S(E;h)$ is then obtained from Eq.~(\ref{eq:H}) with
$E/N = -m^2/2 - hm$. Since $E$ is a quadratic function of $m$, $S(E;h)$
obtained in this way has two branches. In numerical simulations, we
always observe only the upper branch, since it has tremendously 
more number of states than the lower branch. We confirmed that
the numerically obtained $S(E;h)$ is in a good agreement with 
the upper branch of the analytic one. However, since Stirling's 
formula fails at $m=\pm 1$, we instead use the numerically
obtained $S(E;h)$ for the time evolution of the system. 

The entropy $S$ is not a dynamic quantity but a thermodynamic quantity. 
In the present work, we consider the case that the external magnetic field 
$h(t) = h_0 \sin(\Omega t)$ varies with time so slowly.  In this low-frequency 
limit (i.e., the driving frequency $\Omega$ is sufficiently low), 
we may use the adiabatic approximation: 
At a given instant of time $t$, the entropy $S(E; h)$ computed above 
at constant external field $h$ is used for the time evolution of the 
system under the time-dependent field $h(t)$. 
In practice, $S(E; h)$ is computed at 100 different values of $h$ in the range
$0 \leq h \leq h_0$ with $h_0$ set equal to $0.1$ throughout the work, and
for a given value of $h(t)$, we choose $S(E; h)$ with the value of $h$ closest to 
that of $h(t)$. 
From the symmetry of Hamiltonian in Eq.~(\ref{eq:H}), one has $S(E; -h) = S(E; h)$, 
which is utilized for the time evolution during the period $h(t) < 0$.

\begin{figure}
\centering{\resizebox*{!}{12cm}{\includegraphics{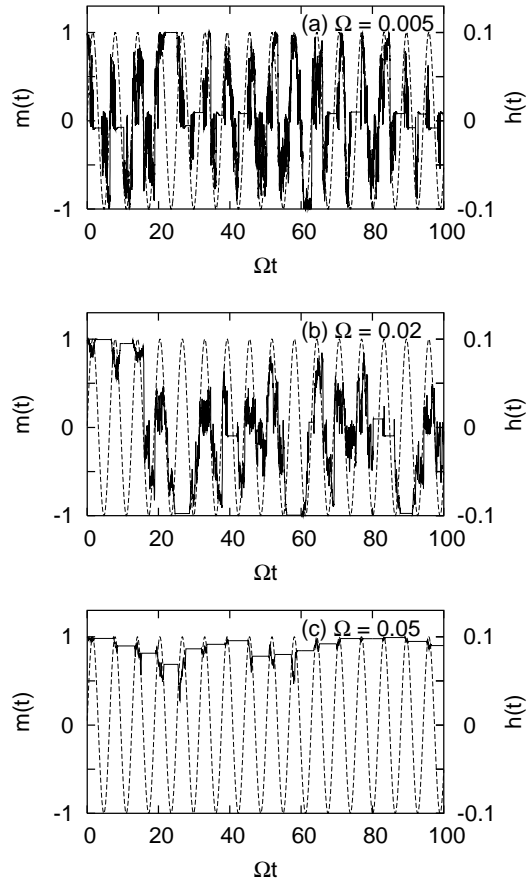}}}
\caption{Entropic sampling time evolution of the magnetization $m(t)$
in a system of size $N=1600$ at driving frequency $\Omega = $ (a) 0.005, (b) 0.02, and
(c) 0.05. Time series data have been obtained from the initial condition
that all spins are aligned to have $m(t=0) = 1$.  For comparison, the time evolution
of the external magnetic field $h(t) = h_0 \sin(\Omega t)$ with $h_0 = 0.1$ is also 
displayed (see the dashed lines). 
At sufficiently small values of $\Omega$, $m(t)$ is observed to follow the external field 
$h(t)$ well. 
As $\Omega$ is raised, such coherence behavior is shown to become weaker. 
}
\label{fig:mtw}
\end{figure}

\section{Results}
The time evolution of the magnetization $m(t) \equiv M(t)/N
= (1/N) \sum_i \sigma_i(t)$ is displayed in Fig.~\ref{fig:mtw}
for the system of size $N=1600$ at driving frequencies 
$\Omega = $ (a) 0.005, (b) 0.02, and (c) 0.05. 
Observed is the coherence behavior at lower frequencies between
$m(t)$ and $h(t)$, which may be explained in the following way:
When $h(t) > 0$, the energy of the system is low for a larger value of $m(t)$.
Since the entropy is a monotonically increasing function of the energy
and since the entropic sampling dynamics tends to decrease the
entropy, the positive value of $h(t)$ drives the system to have a larger value
of $m(t)$, resulting in the coherence behavior observed in Fig.~\ref{fig:mtw}.
At small driving frequencies, the system has enough time to adjust itself 
to follow the external driving, and thus exhibits coherence behavior
between the magnetization and the external driving well. 
On the other hand, as the driving frequency becomes higher, i.e., 
as the external driving changes in time too fast, 
the spins in the system do not have time enough to follow the driving $h(t)$.

\begin{figure}
\centering{\resizebox*{!}{5cm}{\includegraphics{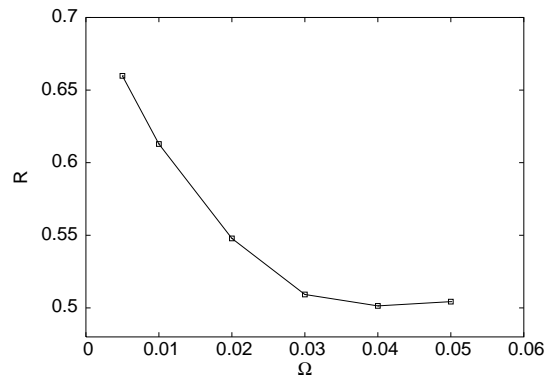}}}
\caption{Occupancy ratio $R$ as a function of the driving frequency $\Omega$ 
for the globally coupled Ising model of size $N=1600$. 
As $\Omega$ becomes higher, coherence is observed to diminish:
$R(\omega) \rightarrow 0.5$ as $\Omega$ is increased. 
}
\label{fig:orw}
\end{figure}

Such coherence behavior in Fig.~\ref{fig:mtw}, which we call {\it entropic coherence}
to emphasize the role of the entropy, can be quantitatively detected
by the occupancy ratio $R$, defined to be the average fraction of the spins in 
the direction of $h(t)$~\cite{bjkim,ORfirst}:
\begin{equation}
R\equiv \left\langle\frac{\mbox{number of spins in the direction of}~h(t)}
{\mbox{total number of spins}}\right\rangle_t 
\label{eq:OR}
\end{equation}
with $\langle\cdots\rangle_t$ denoting the time average.
For the time sequence like the one in Fig.~\ref{fig:mtw}(c), the occupancy ratio $R$ has
a value close to 1/2 since very few spins follow the driving $h(t)$ during the time 
period $h(t) <0$.
The occupancy ratio as a function of the driving frequency 
is displayed in Fig.~\ref{fig:orw} for the system of size $N=1600$.
As expected from Fig.~\ref{fig:mtw}, the occupancy ratio is observed to decrease with $\Omega$
and approaches the value $1/2$, signaling the disappearance of the coherence behavior 
at higher frequencies.

From the above observation, it is expected that the competition
of time scales should play an important role in the coherence behavior.
We use the reasoning similar to that in the standard stochastic resonance~\cite{bjkim} 
and identify the two times scales as follows: One is extrinsic and originates from 
the external driving frequency, and the other is intrinsic, in principle not depending
upon the external driving.  The extrinsic time scale $\tau_{\rm ext} \equiv \Omega^{-1}$ 
reduces as the driving frequency $\Omega$ is increased, while the intrinsic 
time scale $\tau$ is fixed and independent of $\Omega$. 
When $\tau_{\rm ext} \gg \tau$, the system has enough time to relax and to follow $h(t)$, 
while in the opposite limit $\tau_{\rm ext} \ll \tau$, spins cannot follow $h(t)$ 
which varies too fast. 
The intrinsic time scale $\tau$, computed from the correlation
function $\langle m(t) m(0) \rangle$ in equilibrium with the ensemble average
$\langle \cdots \rangle$, has been shown to follow the simple
power law $\tau \sim N$~\cite{soc:mychoi}.  
Qualitatively, one can also interpret $\tau$ as the time scale 
required to overcome the entropic barrier~\cite{entropicbarrier},
separating $m = 1$ and $m = -1$.

\begin{figure}
\centering{\resizebox*{!}{12cm}{\includegraphics{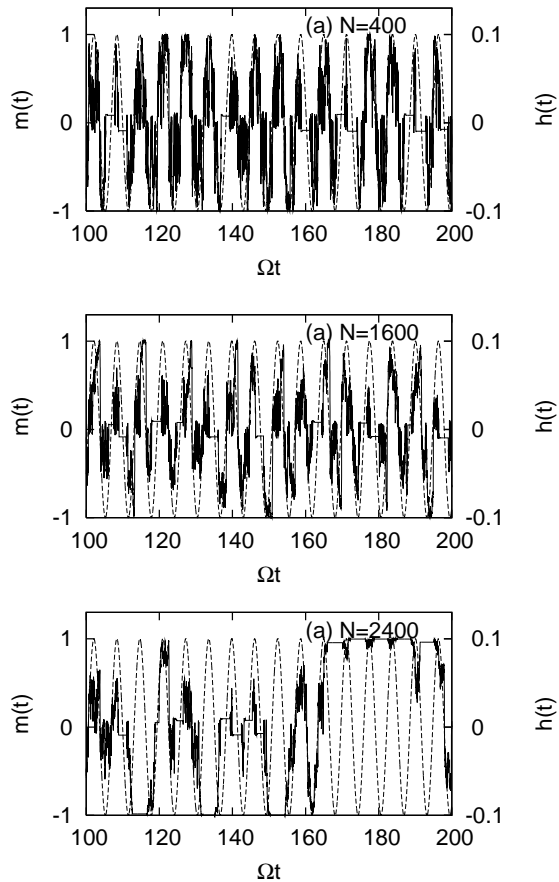}}}
\caption{Entropic sampling time evolution for the globally coupled Ising model
at driving frequency $\Omega = 0.01$ for size $N = $ (a) 400, (b) 1600, and
(c) 2400.  As $N$ is increased, the coherence behavior becomes weaker.
}
\label{fig:mtN}
\end{figure}

\begin{figure}
\centering{\resizebox*{!}{5cm}{\includegraphics{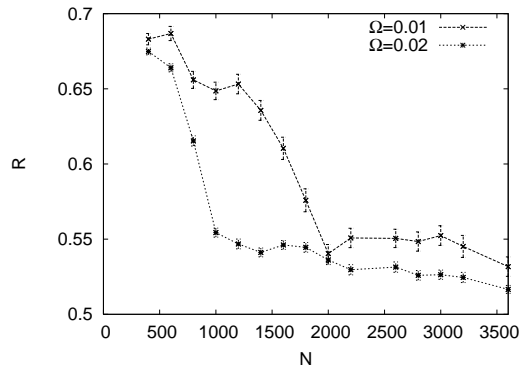}}}
\caption{Occupancy ratio $R$ versus the system size $N$
at driving frequency $\Omega = 0.01$ and 0.02. 
As $N$ is increased, entropic coherence becomes weaker
and gradually disappears in sufficiently large systems.
}
\label{fig:orN}
\end{figure}

Figure~\ref{fig:mtN} shows the time evolution of the magnetization $m(t)$
in the presence of external periodic driving $h(t) = h_0 \sin(\Omega t)$
with $\Omega = 0.01$ for the system size $N=$ (a) 400, (b) 1600, and (c) 2400.
As the size $N$ is increased, the internal time scale $\tau$ grows, and 
eventually when $\tau$ becomes too large compared with $\tau_{\rm ext}$, 
the spins in the system cannot follow the external field. 
The behavior of the occupancy ratio $R$ with the size $N$ is shown 
in Fig.~\ref{fig:orN} at external frequencies $\Omega =0.01$ and 
0.02, which manifests the disappearance of the coherence 
behavior as the system becomes larger~\cite{size}. 
Such a behavior may have interesting implications in sociological systems:
Social collective behavior in accord with external driving may disappear quickly
as the group size exceeds the critical value. 
In other words, the control by means of an external agent (e.g., laws or other
social regulations) may become completely inefficient if the society grows too large. 
When this happens, enhancing interactions among people rather than 
strengthening social enforcement can be more effective for collective behavior to emerge.

\section{Conclusion}
In summary, we have studied the coherence behavior in the 
globally coupled Ising system under a time-periodic 
external magnetic field. The entropic sampling algorithm has been
used both for the calculation of the entropy and for the time
evolution of the system. When the external driving frequency is sufficiently 
low (namely, when the external field varies slowly with time) 
and when the system is not too large, the coherence behavior between the magnetization 
and the external field has been observed. 
It seems appropriate to call this phenomenon entropic coherence, which 
emphasizes the role of the entropy in the time evolution. 
It is noteworthy that if one uses the similar entropic sampling time
evolution for the entropy as a function of, e.g., the magnetization
$S(M)$ instead of the energy, we do not have the coherence behavior: 
$S(M)$ can be computed from a purely combinatorial counting problem and accordingly,
does not reflect the presence of the external magnetic field in the Hamiltonian~(\ref{eq:H}).
It is plausible that this entropic coherence idea may be used to explain the coherence behavior
of those systems the dynamics of which are more appropriately described by the entropy.
The importance of the entropy has been pointed out in the description of 
the biological evolution~\cite{mychoi} and the resulting dynamics has been suggested 
to be of wide applicability to a variety of the self-organized critical systems~\cite{soc:mychoi}.
Accordingly, we believe that the present work can be extended to
explain, e.g., various social and biological systems in the presence of periodic external driving 
coupled to the internal degrees of freedom. 

\ack
This work was supported in part by KOSEF Grant No.\ R14-2002-062-01000-0, 
by the Hwang-Pil-Sang research fund (B.J.K), and by the BK21 Program (M.Y.C.). 
Numerical simulations were performed on the computer cluster Iceberg at Ajou University.


\begin{thebibliography}{99}
\bibitem{size}
The origin of the non-monotonic behavior at small sizes in Fig.~\ref{fig:orN}
is not clear at this stage; it can be either due to  a numerical artifact in the computation 
of the entropy $S(E;h)$, or it may originate from a generalized time-scale matching,
   i.e., the internal time scale, increasing with the system size $N$, 
   matches with an integer multiple of the external time scale.
\end{thebibliography}
\end{document}